\documentclass[12pt,draftclsnofoot,onecolumn]{IEEEtran}

\usepackage{amsfonts}
\usepackage{amsmath}
\usepackage{bm}
\usepackage{mathrsfs}
\usepackage{multirow}
\usepackage{booktabs}
\usepackage{setspace}
\usepackage{amssymb}
\usepackage{mdwmath}
\usepackage{balance}
\usepackage{color}
\usepackage{citesort}
\usepackage{url}
\usepackage{threeparttable}
\usepackage{ifpdf}
\usepackage[dvips]{graphicx}
\graphicspath{{../}}
\DeclareGraphicsExtensions{.pdf,.jpeg,.png,.eps}
\usepackage[tight,footnotesize]{subfigure}

\ifCLASSOPTIONonecolumn

\newcommand{\figwidth}{0.6\textwidth}
\newcommand{\vspacecaption}{\vspace{0.0cm}}
\newcommand{\vspacefigure}{\vspace{0.0cm}}

\fi
\ifCLASSOPTIONtwocolumn

\newcommand{\figwidth}{0.48\textwidth}
\newcommand{\vspacecaption}{\vspace{-0.0cm}}
\newcommand{\vspacefigure}{\vspace{-0.0cm}}

\fi

\begin{document}

\title{Block Markov Superposition Transmission of RUN Codes}

\author{Chulong~Liang,
        Xiao~Ma,~\IEEEmembership{Member,~IEEE,}
        and~Baoming~Bai,~\IEEEmembership{Member,~IEEE}
\thanks{Generated on \today.}
\thanks{This work was supported in part by the 973 Program under Grant 2012CB316100 and in part by the National Natural Science Foundation of China under Grant 91438101.}
\thanks{Chulong~Liang was with the Department of Electronics and Communication Engineering, Sun Yat-sen University, Guangzhou 510006, China. He is now with the Department of Electronic Engineering, City University of Hong Kong, Kowloon, Hong Kong (e-mail: lchul@mail2.sysu.edu.cn).}
\thanks{Xiao~Ma is with the Department of Electronics and Communication Engineering, Sun Yat-sen University, Guangzhou 510006, China (e-mail: maxiao@mail.sysu.edu.cn).}
\thanks{Baoming~Bai is with the State Key Laboratory of Integrated Services Networks, Xidian University, Xi'an 710071, Shaanxi, China (e-mail: bmbai@mail.xidian.edu.cn).}
}

\maketitle

\begin{abstract}
In this paper, we propose a simple procedure to construct (decodable) good codes with any given alphabet (of moderate size) for any given (rational) code rate to achieve any given target error performance (of interest) over additive white Gaussian noise (AWGN) channels. We start with constructing codes over groups for any given code rates. This can be done in an extremely simple way if we ignore the error performance requirement for the time being. Actually, this can be satisfied by repetition (R) codes and uncoded~(UN) transmission along with time-sharing technique. The resulting codes are simply referred to as RUN codes for convenience. The encoding/decoding algorithms for RUN codes are almost trivial. In addition, the performance can be easily analyzed. It is not difficult to imagine that a RUN code usually performs far away from the corresponding Shannon limit. Fortunately, the performance can be improved as required by spatially coupling the RUN codes via block Markov superposition transmission~(BMST), resulting in the BMST-RUN codes. Simulation results show that the BMST-RUN codes perform well (within one dB away from Shannon limits) for a wide range of code rates {\color{black}and outperform the BMST with bit-interleaved coded modulation~(BMST-BICM) scheme.}
\end{abstract}


\section{Introduction}\label{sec:Introduction}

Since the invention of turbo codes~\cite{Berrou93} and the rediscovery of low-density parity-check~(LDPC) codes~\cite{Gallager63}, many turbo/LDPC-like codes have been proposed in the past two decades.
Among them, the convolutional LDPC codes~\cite{Felstrom99}, recast as spatially coupled LDPC~{\color{black}(SC-LDPC)} codes in~\cite{Kudekar11}, exhibit a threshold saturation phenomenon and were proved to have better performance than their block counterparts.
In a certain sense, the terminology ``spatial coupling" is more general, as can be interpreted as making connections among independent subgraphs, or equivalently, as introducing memory among successive independent transmissions.
With this interpretation, braided block codes~\cite{Feltstrom09} and staircase codes~\cite{Smith12}, as the convolutional versions of (generalized) product codes, can be classified as spatially coupled codes. In~\cite{Moloudi14}, the spatially coupled version of turbo codes was proposed, whose belief propagation~(BP) threshold is also better than that of the uncoupled ensemble.

{\color{black}Recently, block Markov superposition transmission~(BMST)~\cite{Ma13,Ma15,Liang14c}
was proposed}, which can also be viewed as the spatial coupling of generator matrices of short codes.
{\color{black}The original BMST codes are defined over the binary field $\mathbb{F}_2$.
In~\cite{Ma15}, it has been pointed out that any code with fast encoding algorithms and soft-in soft-out~(SISO) decoding algorithms can be taken as the basic code.
For example, one can take the Hadamard transform~(HT) coset
codes as the basic codes, resulting in a class of multiple-rate codes with rates ranging from $1/2^p$ to $(2^p-1)/2^p$, where $p$ is a positive integer~\cite{Hu14,Liang15}.
Even more flexibly, one can use the repetition and/or single-parity-check~(RSPC) codes as the basic codes to construct a class of multiple-rate codes with rates ranging from $1/N$ to $(N-1)/N$, where $N>1$ is an integer~\cite{Hu15}.
It has been verified by simulation that the construction approach is applicable not only to binary phase-shift keying~(BPSK) modulation  but also to bit-interleaved coded modulation~(BICM)~\cite{Liang14}, spatial modulation~\cite{Yang14}, continuous phase modulation~(CPM)~\cite{Liu15}, and intensity modulation in visible light communications~(VLC)~\cite{Xu15}.}

In this paper, we propose a procedure to construct codes over groups, which extends the construction of BMST-RSPC codes~\cite{Hu15} in the following two aspects.
First, we allow uncoded symbols occurring in the basic codes. Hence the encoding/decoding algorithms for the basic codes become simpler.
Second, we derive a performance union bound for the repetition codes with any given signal mapping, which is critical for designing good BMST codes without invoking simulations.
We will not argue that the BMST construction can always deliver better codes than other existing constructions.\footnote{Actually, compared with SC-LDPC codes, the BMST codes usually have a higher error floor.
However, the existence of the high error floor is not a big issue since it can be lowered if necessary by increasing the encoding memory.
}
Rather, we argue that the proposed one is more flexible in the sense that it applies to {\em any} given signal set~(of moderate size), {\em any} given~(rational) code rate and {\em any} target error performance~(of interest).
We start with constructing group codes, referred to as RUN codes, with any given rate by time-sharing between repetition~(R) codes and/or uncoded~(UN) transmission.
By transmitting the RUN codes in the BMST manner, we can have a class of good codes~(called BMST-RUN codes). The performance of a BMST-RUN code is closely related to the encoding memory and can be predicted analytically in the high signal-to-noise~ratio~(SNR) region with the aid of the readily-derived union bound.
Simulation results show that the BMST-RUN codes can approach the Shannon limits at any given target error rate (of interest) in a wide range of code rates over
\textcolor{black}{both} additive white Gaussian noise~(AWGN) channels {\color{black}and Rayleigh flat fading channels}.

{\color{black}The pragmatic reader may question the necessity to construct codes over high-order signal constellations, since bandwidth efficiency can also be attained by BICM with binary codes. However, in addition to the flexility of the construction, the BMST-RUN codes have the following competitive advantages.
\begin{itemize}
  \item BMST-RUN codes can be easily designed to obtain shaping gain in at least two ways. One is designing codes directly over a well-shaped signal constellation, say, non-uniformly spaced constellation~\cite{Sun93}. The other is implementing Gallager mapping for conventional signal constellations~\cite{Ma04}.
      In both cases, neither optimization for bit-mapping~(at the transmitter) nor iterations between decoding and demapping~(at the receiver) are required.
  \item BMST-RUN codes can be defined over signal sets of any size, such as 3-ary pulse amplitude modulation~(3-PAM) and 5-PAM, which can be useful to transmit real samples directly~\cite{Yang12}.
\end{itemize}
}

{\color{black}The rest of this paper is organized as follows. In Section~\ref{sec:ReviewOfBMST}, we take a brief review of the BMST technique. In Section~\ref{sec:CodesOverGroups}, we discuss constructing group codes with any given signal set and any given code rate. In Section~\ref{sec:BMSToverGroups}, we propose the construction method of BMST-RUN codes and discuss the performance lower bound. In Section~\ref{sec:Examples}, we give simulation results and make a performance comparison between the BMST-RUN codes and the BMST-BICM scheme. In Section~\ref{sec:Conclusion}, we conclude this paper.}

{\color{black}
\section{Review of Binary BMST Codes}\label{sec:ReviewOfBMST}
Binary BMST codes are convolutional codes with large constraint lengths~\cite{Ma13,Ma15}. Typically,
a binary BMST code of memory $m$ consists of a short code~(called the \emph{basic code}) and at most $m+1$ interleavers~\cite{Liang14c}. Let $\mathcal{C}[n,k]$ be the basic code defined by a $k \times n$ generator matrix $\bm{G}$ over the binary field $\mathbb{F}_2$. Denote $\bm{u}^{(0)}, \bm{u}^{(1)}, \cdots, \bm{u}^{(L-1)}$ as $L$ blocks of data to be transmitted, where $\bm{u}^{(t)} \in \mathbb{F}^k_2$ for $0 \leq t \leq L-1$. Then, the encoding output $\bm{c}^{(t)} \in \mathbb{F}^n_2$ at time $t$ can be expressed as~\cite{Liang14c}
\begin{equation}
\bm{c}^{(t)} = \bm{u}^{(t)}\bm{G}\bm{\varPi}_0 + \bm{u}^{(t-1)}\bm{G}\bm{\varPi}_1 + \cdots + \bm{u}^{(t-m)}\bm{G}\bm{\varPi}_m,
\end{equation}
where $\bm{u}^{(t)}$ is initialized to be $\mathbf{0} \in \mathbb{F}^k_2$ for $t<0$ and $\bm{\varPi}_0, \cdots, \bm{\varPi}_m$ are $m+1$ permutation matrices of order $n$.
For $L \leq t \leq L+m-1$, the zero message sequence $\bm{u}^{(t)} = \mathbf{0} \in \mathbb{F}^k_2$ is input into the encoder for termination.
Then, $\bm{c}^{(t)}$ is mapped to a signal vector $\bm{s}^{(t)}$ and transmitted over the channel, resulting in a received vector $\bm{y}^{(t)}$.

At the receiver, the decoder executes the sliding-window decoding~(SWD) algorithm to recover the transmitted data~$\bm{u}^{(0)}, \cdots, \bm{u}^{(L-1)}$~\cite{Ma13,Ma15}. Specifically, for an SWD algorithm with a decoding delay $d$, the decoder takes $\bm{y}^{(t)}, \cdots, \bm{y}^{(t+d)}$ as inputs to recover $\bm{u}^{(t)}$ at time $t+d$, which is similar to the window decoding~(WD) of the SC-LDPC codes~\cite{Lentmaier10,Iyengar12,Iyengar13}.
The structure of the BMST codes also admits a two-phase decoding (TPD) algorithm~\cite{Liang14c}, which can be used to reduce the decoding delay and to predict the performance in the extremely low bit-error-rate~(BER) region.

As discussed in~\cite{Ma15}, binary BMST codes have the following two attractive features.
\begin{enumerate}
  \item Any code~(linear or nonlinear) can be the basic code as long as it has fast encoding algorithms and SISO decoding algorithms.

  \item Binary BMST codes have a simple genie-aided lower bound when transmitted over AWGN channels using BPSK modulation, which shows that the maximum extra coding gain can approach $10\log_{10}(m+1)$~dB compared with the basic code. {\color{black}The tightness of this simple lower bound in the high SNR region under the SWD algorithm has been verified by both the simulation and the extrinsic information transfer~(EXIT) chart analysis~\cite{Huang15}}.
\end{enumerate}

Based on the above two facts, a general procedure has been proposed for constructing capacity-approaching codes at any
given target error rate~\cite{Liang14c}.
Suppose that we want to construct a binary BMST code of rate $R$ at a target BER of $p_{\rm target}$. First, we find a rate-$R$ short code~$\mathcal{C}$ as the basic code. Then, we can determine the encoding memory $m$ by \begin{equation}\label{eq:ComputeMemory}
        m = \left\lceil 10^{\frac{\gamma_{\rm target} - \gamma_{\lim}}{10}}-1 \right\rceil,
        \end{equation}
where $\gamma_{\rm target}$ is the minimum SNR for the code $\mathcal{C}$ to achieve the BER $p_{\rm target}$,  $\gamma_{\lim}$ is the Shannon limit corresponding to the rate $R$, and $\left\lceil x \right\rceil$ stands for the minimum integer greater than or equal to $x$. Finally, by generating $m+1$ interleavers uniformly at random, the BMST code is constructed.
With this method, we have constructed a binary BMST code of memory $30$ using the Cartesian product of the R code $[2,1]^{5000}$, which has a predicted BER lower than $10^{-15}$ within one dB away from the Shannon limit.

\section{RUN Codes over Groups}\label{sec:CodesOverGroups}
\subsection{System Model and Notations}
Consider a symbol set $\mathcal{M} = \{0, 1, \cdots, q-1 \}$ and an $\ell$-dimensional signal constellation $\mathcal{A} \subset \mathbb{R}^\ell$ of size $q$. The symbol set $\mathcal{M}$ can be treated as a group by defining the operation $u \oplus w = (u+w) \mod q$ for $u, w \in \mathcal{M}$ . Let $\varphi$ be a (fixed) one-to-one mapping $\varphi: \mathcal{M} \rightarrow \mathcal{A}$. Let $u \in \mathcal{M}$ be a symbol to be transmitted.
For the convenience of performance analysis, instead of transmitting $\varphi(u)$ directly, we transmit a signal $s = \varphi(u \oplus w)$, where $w$ is a sample of a uniformly distributed random variable over $\mathcal{M}$ and assumed to be known at the receiver.
The received signal $y = s + z$, where $+$ denotes the component-wise addition over $\mathbb{R}^\ell$ and $z$ is an $\ell$-dimensional sample from
a zero-mean white Gaussian noise process with variance~$\sigma^2$ per dimension.
The SNR is defined as
\begin{equation}\label{eq:uncoded}
{\rm SNR} = \frac{\sum_{s\in\mathcal{A}}\|s\|^2}{\ell\sigma^2q},
\end{equation}
where $\|s\|^2$ is the squared Euclidean norm of $s$.

In this paper, for a discrete random variable $V$ over a finite set $\mathcal{V}$, we denote its {\em a priori message} and {\em extrinsic message} as $P^a_{V}(v), v \in \mathcal{V}$ and $P^e_{V}(v), v \in \mathcal{V}$, respectively. A SISO decoding is a process that takes {\em a priori} messages as inputs and delivers extrinsic messages as outputs. We assume that the information messages are independent and uniformly distributed~(i.u.d.) over $\mathcal{M}$.

\subsection{Repetition~(R) Codes}\label{subsec:SystemModel}
\begin{figure}[t]
\centering
  \includegraphics[width=\figwidth]{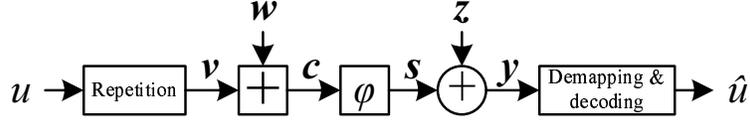}\\
  \vspacecaption
  \caption{\color{black}A message $u$ is encoded into $\bm{v} = (u,\cdots,u)$ and transmitted over AWGN channels.}
  \label{fig:SystemModel}
  \vspacefigure
\end{figure}
{\color{black}Fig.~\ref{fig:SystemModel} shows the transmission of a message $u$ for $N$ times over AWGN channels.}

\subsubsection{Encoding} The encoder of an R code $\mathcal{C}[N, 1]$ over $\mathcal{M}$ takes as input a single symbol $u \in \mathcal{M}$ and delivers as output an $N$-dimensional vector $\bm{v} = \left( v_0, \cdots, v_{N-1} \right) = \left( u, \cdots, u \right)$.

\subsubsection{Mapping} The $j$-th component $v_j$ of the codeword $\bm{v}$ is mapped to the signal $s_j = \varphi( v_j {\oplus} w_j )$ for $j=0,\cdots,N-1$, where $\bm{w} = (w_0,\cdots,w_{N-1})$ is a random vector sampled from an i.u.d. process over $\mathcal{M}$.

\subsubsection{Demapping} Let $\bm{y} = (y_0, \cdots, y_{N-1})$ be the received signal vector corresponding
to the codeword $\bm{v}$. The {\em a priori} messages input to the decoder are computed as
\begin{equation}\label{eq:channelAPP}
P^a_{V_j}\left(v\right) \propto \exp\left(-\frac{\|y_j-\varphi(v \oplus w_j)\|^2}{2\sigma^2} \right), v \in \mathcal{M}
\end{equation}
for $j=0,\cdots,N-1$.

\subsubsection{Decoding}
The SISO decoding algorithm computes the {\em a posteriori} messages
\begin{equation}
P^e_{U}(u) \propto \prod_{0 \leq \ell \leq N-1} P^a_{V_\ell}(u), u \in \mathcal{M}
\end{equation}
for making decisions and the extrinsic messages
\begin{equation}\label{eq:RUN_SISO_decoding}
P^e_{V_j}(v) \propto \prod_{0 \leq \ell \leq N-1, \ell \neq j} P^a_{V_\ell}(v), v \in \mathcal{M}
\end{equation}
for $j=0,\cdots,N-1$ for iteratively decoding when coupled with other sub-systems.

\subsubsection{Complexity} Both the encoding/mapping and the demapping/decoding have linear computational complexity per coded symbol.

\subsubsection{Performance} Let $\hat{u}$ denote the hard decision output. The performance is measured by the symbol-error-rate~(SER) ${\rm SER} \triangleq \Pr\{ \hat{U} \neq U \} = \sum_{u \in \mathcal{M}} \frac{1}{q} \Pr\{ \hat{U} \neq U | U = u \}$. Define $e=\hat{u} \ominus u$, where $\ominus$ denotes the subtraction under modulo-$q$ operation. Due to the existence of the random vector $\bm{w}$, the peformance is irrelevant to the transmitted symbol $u$. We define
\begin{equation}\label{eq:DeX}
D_{e}\left( X \right) = \sum_{w \in \mathcal{M}} \frac{1}{q} X^{\|\varphi(w) - \varphi(e \oplus w)\|^2}
\end{equation}
as the average Euclidian distance enumerating function~(EDEF) corresponding to the error $e$, where $X$ is a dummy variable. Then, the average EDEF $B^{(N)}\left(X\right)$ for the R code $\mathcal{C}[N,1]$ over all possible messages $u$ and all possible vectors $\bm{w}$ can be computed as
\begin{align}
&B^{(N)}(X) \nonumber \\
&= \sum_{e \in \mathcal{M}} \sum_{\bm{w} \in \mathcal{M}^N} \frac{1}{q^N} \sum_{u \in \mathcal{M} } \frac{1}{q} X^{\sum\limits_{j=0}^{N-1}\|\varphi(u \oplus w_j) - \varphi(u \oplus e \oplus w_j)\|^2} \nonumber \\
&= \sum_{e \in \mathcal{M}} (D_{e}(X))^N \triangleq \sum_{\delta} B_{\delta}^{(N)}X^{\delta^2},
\end{align}
where $B_{\delta}^{(N)}$ denotes the average number of signal pairs $(\bm{s}, \hat{\bm{s}})$ with Euclidean distance $\delta$, $\bm{s} = \left( \varphi(u \oplus w_0), \cdots, \varphi(u \oplus w_{N-1}) \right)$ and $\hat{\bm{s}} = \left( \varphi(\hat{u} \oplus w_0), \cdots, \varphi(\hat{u} \oplus w_{N-1}) \right)$. The performance {\color{black}under the mapping $\varphi$} can be upper-bounded by the union bound as
\begin{equation}\label{eq:RcodeUnionBound}
{\rm SER} = f_{\varphi,N} ({\rm SNR}) \leq \sum_{\delta > 0} B^{(N)}_{\delta}{\rm Q}\left( \frac{\delta}{2\sigma} \right),
\end{equation}
where ${\rm Q}\left( \frac{\delta}{2\sigma} \right)$ is the pair-wise error probability with ${\rm Q}\left(x\right)\triangleq\int_{x}^{+\infty}\frac{1}{\sqrt{2\pi}}\exp\left(-\frac{z^2}{2}\right)dz$.

\begin{figure}[t]
\centering
  \includegraphics[width=\figwidth]{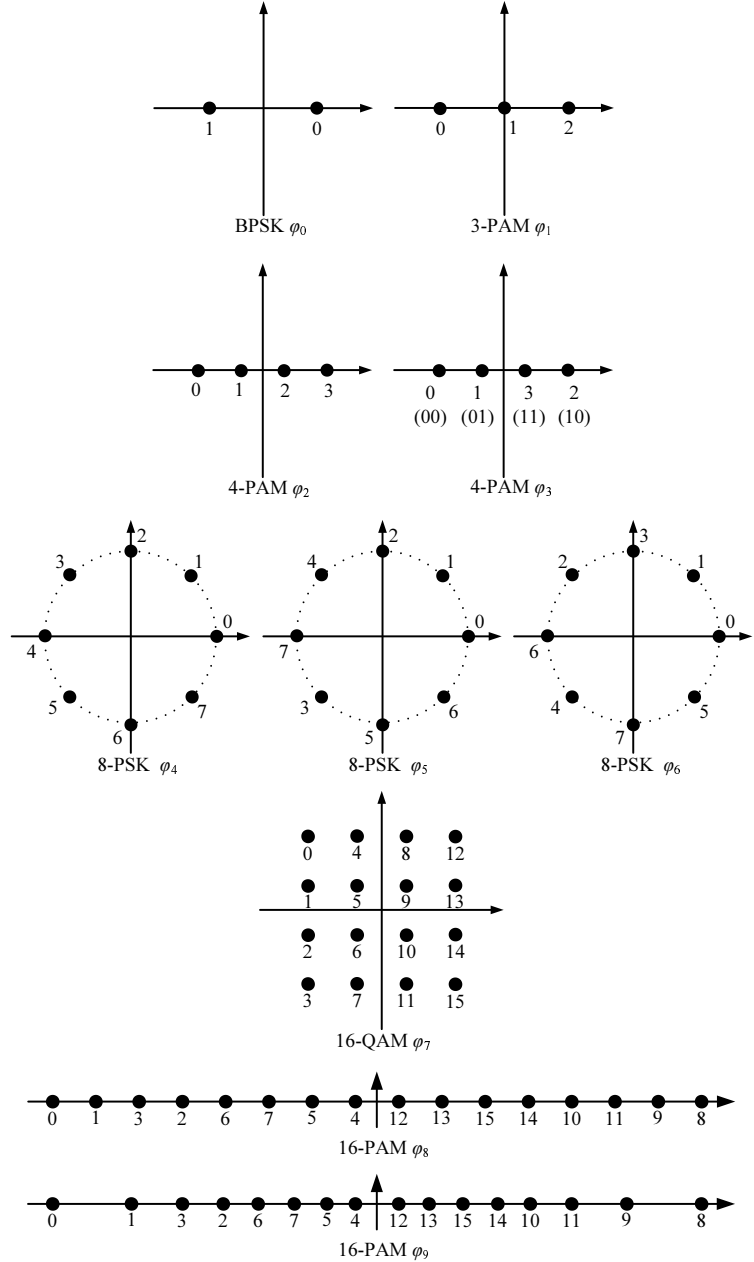}\\
  \vspacecaption
  \caption{Examples of signal constellations and mappings.}
  \label{fig:SignalSetsAndMappings}
  \vspacefigure
\end{figure}
\begin{figure}[t]
\centering
  \includegraphics[width=\figwidth]{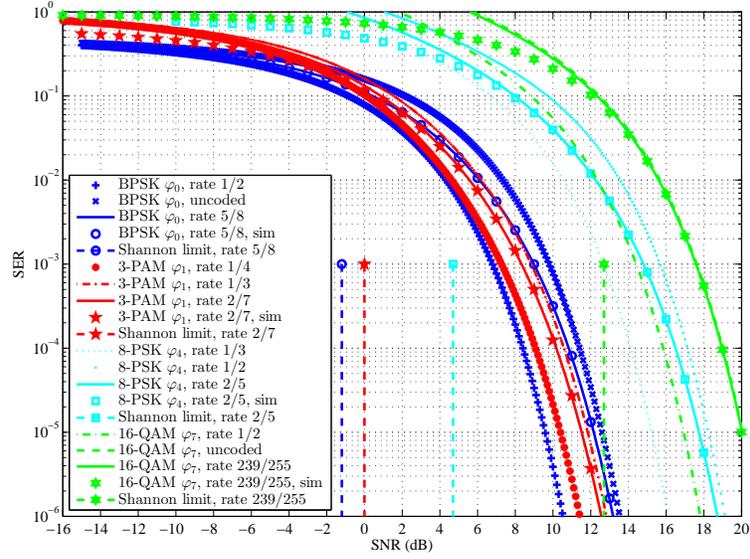}\\
  \vspacecaption
  \caption{Performances and bounds of RUN codes. The ``rate'' in the legend of this figure~(or other similar figures in this paper) refers to the code rate. A rate-$R$ code over a $q$-ary constellation has a spectral efficiency of $R \log_2(q)$ in bits per symbol, at which the Shannon limit is determined.}
  \label{fig:PerformanceRUNcodes}
  \vspacefigure
\end{figure}
\begin{figure}[t]
\centering
  \includegraphics[width=\figwidth]{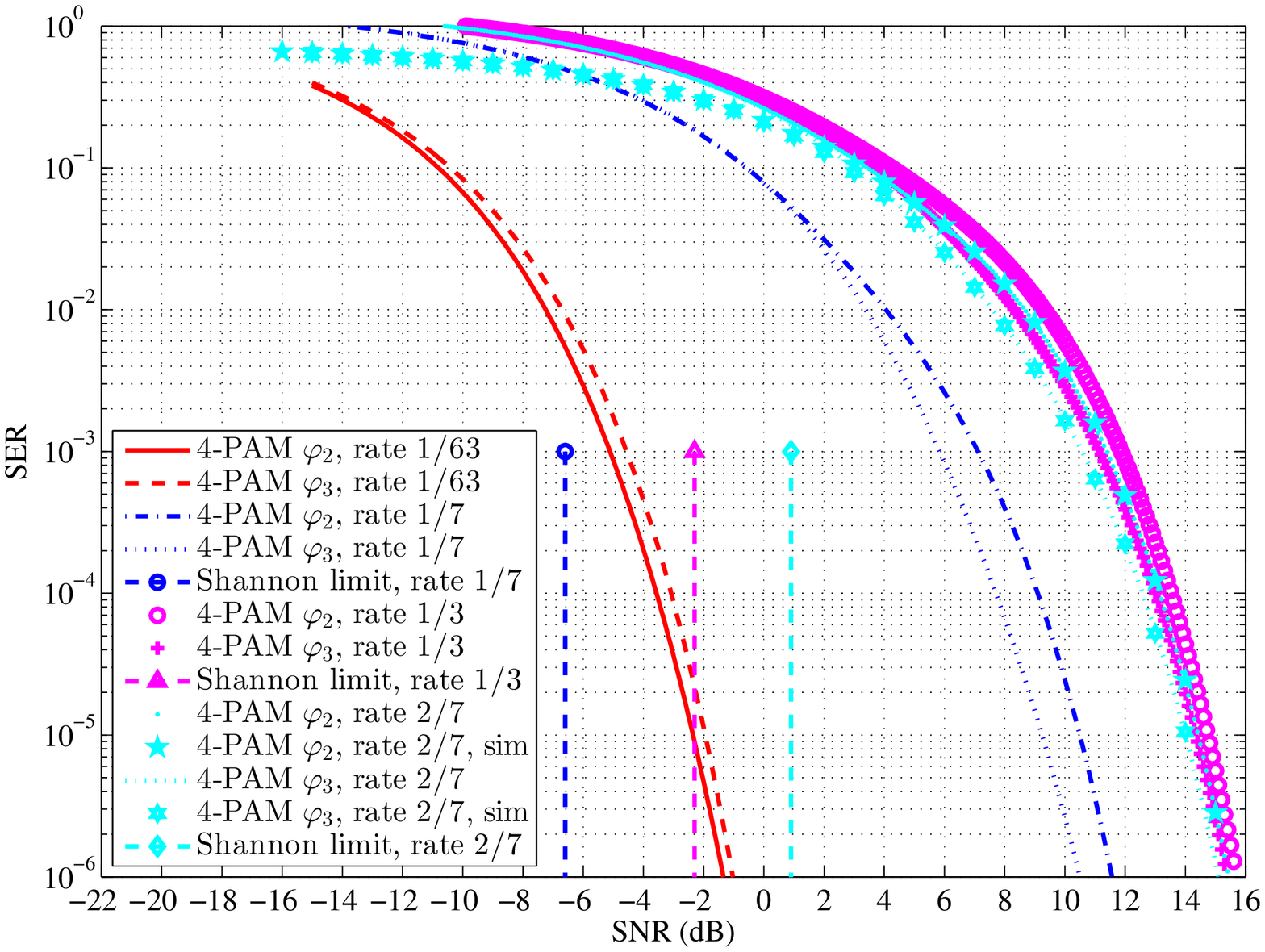}\\
  \vspacecaption
  \caption{Performances and bounds of R codes with 4-PAM under different mappings.}
  \label{fig:DifferentMapping}
  \vspacefigure
\end{figure}
{\color{black}
From the above derivation, we can see that the performance bounds of the R codes are related to the mapping $\varphi$.
In this paper, we consider as examples the BPSK, the signal set $\{-1,0,+1\}$~(denoted as $3$-PAM), $4$-PAM, $8$-ary phase-shift keying ($8$-PSK) modulation, $16$-ary quadrature amplitude modulation~($16$-QAM), or $16$-PAM, which are depicted in Fig.~\ref{fig:SignalSetsAndMappings} along with mappings denoted by $\varphi_0, \cdots, \varphi_9$ as specified in the figure.
Fig.~\ref{fig:PerformanceRUNcodes} and Fig.~\ref{fig:DifferentMapping} show performance bounds for several R codes defined with the considered constellations.
From the figures, we have the following observations.
\begin{enumerate}
  \item The performance gap between the code $\mathcal{C}[N,1]$ and the uncoded transmission, when measured by the SNR instead of $E_b/N_0$, is roughly $10\log_{10}(N)$~dB.
  \item 
      Given a signal constellation, mappings that are universally good for all R codes may not exist.
      For example, as shown in Fig.~\ref{fig:DifferentMapping}, $\varphi_2$ is better than $\varphi_3$ for rate $1/63$~($N=63$) but becomes worse for rate $1/7$~($N=7$).
\end{enumerate}
}

\subsection{Time-Sharing}\label{subsecTimeSharing}
With repetition codes over groups, we are able to implement code rates $\frac{1}{N}$ for any given integer $N \geq 1$. To implement other code rates, we turn to the time-sharing technique.
To be precise, let $R = \frac{P}{Q}$ be the target rate. There must exist a unique $N \geq 1$ such that $\frac{1}{N+1} < \frac{P}{Q} \leq \frac{1}{N}$.
Then, we can implement a code by time-sharing between the code $\mathcal{C}[N+1, 1]$ and the code $\mathcal{C}[N, 1]$, which is equivalent to encoding $\alpha P$ information symbols with the code $\mathcal{C}[N+1, 1]$ and the remaining $(1-\alpha)P$ symbols with the code $\mathcal{C}[N, 1]$, where $\alpha = \frac{1}{R}-N$ is the time-sharing factor.
Apparently, to construct codes with rate $R>\frac{1}{2}$, we need time-sharing between the code $\mathcal{C}[2,1]$ and the uncoded transmission.
For this reason, we call this class of codes as \emph{RUN codes}, which consist of the R codes and codes obtained by time-sharing between the R codes and/or the uncoded transmission.
We denote a RUN code of rate $\frac{P}{Q}$ as $\mathcal{C}_{\rm RUN}[Q, P]$.
Replacing in Fig.~\ref{fig:SystemModel} the R codes with the RUN codes, we then have a coding system that can transmit messages with any given code rate over any given signal set.
\subsubsection{Encoding} Let $\bm{u} \in \mathcal{M}^P$ be the message sequence. The encoder of the code $\mathcal{C}_{\rm RUN}[Q, P]$ encodes the left-most $\alpha P$ symbols of $\bm{u}$ into $\alpha P$ codewords of $\mathcal{C}[N+1, 1]$ and the remaining symbols into $(1-\alpha)P$ codewords of $\mathcal{C}[N, 1]$.
\subsubsection{Decoding} The decoding is equivalent to decoding separately $\alpha P$ codewords of $\mathcal{C}[N+1, 1]$ and $(1-\alpha)P$ codewords of $\mathcal{C}[N, 1]$.
\subsubsection{Complexity} Both the encoding/mapping and the demapping/decoding have the same complexity as the R codes.
\subsubsection{Performance} The performance of the RUN code of rate $R=\frac{P}{Q}$ is given by
\begin{equation}
{\rm SER} = \alpha \cdot f_{\varphi,N+1}\left( {\rm SNR} \right) + (1-\alpha) \cdot f_{\varphi,N}\left( {\rm SNR} \right),
\end{equation}
which can be upper-bounded with the aid of (\ref{eq:RcodeUnionBound}).
Performances and bounds of several RUN codes defined with
{\color{black}BPSK modulation, $3$-PAM, $4$-PAM, $8$-PSK modulation, or $16$-QAM}
are shown in Fig.~\ref{fig:PerformanceRUNcodes} and Fig.~\ref{fig:DifferentMapping}. We notice that the union bounds with BPSK modulation are the exact performances, while those with other signal sets are upper bounds to the performances. We also notice that the upper bounds become tight as the SER is lower than $10^{-2}$ for all other signal sets.
{\color{black}Not surprisingly, the performances of the RUN codes are far away from the corresponding Shannon limits~(more than $5$~dB) at the SER lower than $10^{-2}$.
}

\section{BMST over Groups}\label{sec:BMSToverGroups}
\subsection{BMST Codes with RUN Codes As Basic Codes}
\begin{figure}[t]
   \centering
   \includegraphics[width=\figwidth]{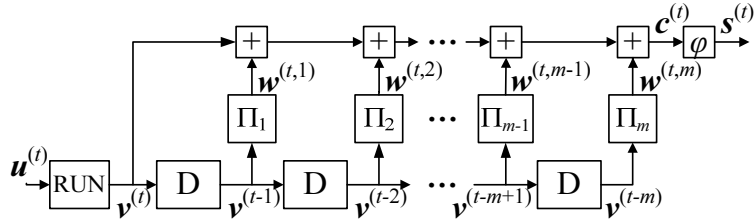}
   \vspacecaption
   \caption{Encoding structure of a BMST-RUN code with memory $m$.}
   \label{fig:encoder}
   \vspacefigure
\end{figure}
We have constructed a class of codes called RUN codes with any given code rate over groups. However, the RUN codes perform far away from the Shannon limits, as evidenced by the examples in Fig.~\ref{fig:PerformanceRUNcodes} {\color{black}and Fig.~\ref{fig:DifferentMapping}}. To remedy this, we transmit the RUN codes in the BMST manner as inspired by the fact that, as pointed out in~\cite{Ma15}, any short code can be embedded into the BMST system to obtain extra coding gain in the low error-rate region. The resulted codes are referred to as BMST-RUN codes. More precisely, we use the $B$-fold Cartesian product of the RUN code $\mathcal{C}_{\rm RUN}[Q,P]$~(denoted as $\mathcal{C}_{\rm RUN}[Q,P]^B$) as the basic code. Fig.~\ref{fig:encoder} shows the encoding structure of a BMST-RUN code with memory $m$, where \fbox{\small RUN} represents the basic encoder, \fbox{$\Pi_1$}, $\cdots$, \fbox{$\Pi_m$} represents $m$ symbol-wise interleavers, \fbox{$+$} represents the superposition with modulo-$q$ addition, and \fbox{$\varphi$} represents the mapping $\varphi$.
Let $\bm{u}^{(t)} \in \mathcal{M}^{PB}$ and $\bm{v}^{(t)} \in \mathcal{M}^{QB}$ be the information sequence and the corresponding codeword of the code $\mathcal{C}_{\rm RUN}[Q,P]^B$ at time $t$, respectively. Then, the sub-codeword $\bm{c}^{(t)}$ can be expressed as
\begin{equation}\label{eq:BMSTRUNencoding}
\bm{c}^{(t)} = \bm{v}^{(t)} \oplus \bm{w}^{(t,1)} \oplus \cdots \oplus \bm{w}^{(t,m)},
\end{equation}
where $\oplus$ denotes the symbol-wise modulo-$q$ addition, $\bm{v}^{(t)} = \mathbf{0} \in \mathcal{M}^{QB}$ for $t<0$ and $\bm{w}^{(t,i)}$ is the interleaved version of $\bm{v}^{(t-i)}$ by the $i$-th interleaver $\Pi_{i}$ for $i=1,\cdots,m$. Then, $\bm{c}^{(t)}$ is mapped to the signal vector $\bm{s}^{(t)} \in \mathcal{A}^{QB}$ symbol-by-symbol and input to the channel. After every $L$ sub-blocks of information sequence, we terminate the encoding by inputting $m$ all-zero sequences $\bm{u}^{(t)}=\mathbf{0} \in \mathcal{M}^{PB}(L \leq t \leq L+m-1)$ to the encoder. The termination will cause a code rate loss. However, the rate loss can be negligible as $L$ is large enough.

\subsection{Choice of Encoding Memory}\label{subsec:BMSToverGroups}

The critical parameter for BMST-RUN codes to approach the Shannon limits at a given target SER is the encoding memory $m$, which can be determined by the genie-aided lower bound.
Essentially the same as for the binary BMST codes~\cite{Ma15}, the genie-aided bound for a BMST-RUN code can be easily derived by assuming all but one sub-blocks $\left\{ \bm{u}^{(i)}, 0 \leq i \leq L-1, i \neq t \right\}$ are known at the receiver. With this assumption, the genie-aided system becomes an equivalent system that transmits the basic RUN codeword $m+1$ times. Hence the performance of the genie-aided system is the same as the RUN code obtained by time-sharing between the code $\mathcal{C}[(N+1)(m+1),1]$ and the code $\mathcal{C}[N(m+1),1]$.
As a result, the genie-aided bound {\color{black}under a mapping $\varphi$} is given by
\begin{equation}
\begin{aligned}
&{\rm SER} = f_{\rm \scriptstyle BMST-RUN}({\rm SNR}, m) \geq f_{\rm \scriptstyle genie}({\rm SNR}, m)& \\
&= \alpha \!\cdot\! f_{\varphi,(N\!+\!1)(m\!+\!1)}\left( {\rm SNR} \right) \!+ \! (1\!-\!\alpha) \!\cdot\! f_{\varphi,N(m\!+\!1)}\left( {\rm SNR} \right),&
\end{aligned}
\end{equation}
which can be approximated using the union bound in the high SNR region.

{\color{black}
Given a signal set $\mathcal{A}$ of size $q$ with labeling $\varphi$, a rate $R=P/Q$ and a target SER $p_{\rm target}$, we can construct a good BMST-RUN code using the following steps.}
\begin{enumerate}
  \item {\color{black}Construct the code $\mathcal{C}_{\rm RUN}[Q,P]^B$ over the modulo-$q$ group by finding $N$ such that $\frac{1}{N+1} < \frac{P}{Q} \leq \frac{1}{N}$ and determining the time-sharing factor $\alpha$ between the R code~$[N+1,1]$ and the R code~$[N,1]$.
      To approach the Shannon limit and to avoid error propagation, we usually choose $B$ such that $QB \geq 1000$.}
  \item Find the Shannon limit $\gamma_{\lim}$ under the signal set $\mathcal{A}$ corresponding to the rate $R$.
  \item \label{step:chooseMemory}{\color{black}Find an encoding memory $m$ such that $m$ is the minimum integer satisfying $f_{\rm \scriptstyle genie}(\gamma_{\lim}, m) \leq p_{\rm target}$.}


  \item Generate $m$ interleavers of size $QB$ uniformly at random.
\end{enumerate}



\subsection{Decoding of BMST-RUN Codes}
\begin{figure}[t]
   \centering
   \includegraphics[width=\figwidth]{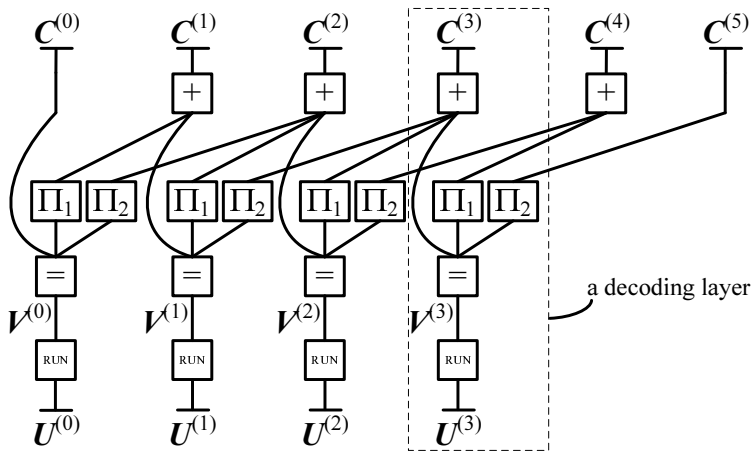}
   \vspacecaption
   \caption{The unified (high-level) normal graph of a BMST-RUN code with $L=4$ and $m=2$.}
   \label{fig:decoder}
   \vspacefigure
\end{figure}
A BMST-RUN code can be decoded by an SWD algorithm with a decoding delay $d$ over its normal graph, which is similar to that of the binary BMST codes~\cite{Ma15}. Fig.~\ref{fig:decoder} shows the unified (high-level) normal graph of a BMST-RUN code with $L=4$ and $m=2$. {The normal graph can also be divided into \emph{layers}{\color{black}, each of which consists of four types of nodes. These nodes represent similar constraints to those for binary BMST codes and have similar message processing as outlined below.}
\begin{itemize}
  \item The process at the node \fbox{\small RUN} is the SISO decoding of the RUN codes as described in Section~\ref{subsec:SystemModel}.
  \item The process at the node \fbox{$=$} can be implemented in the same way as the message processing at a generic variable node of an LDPC code (binary or non-binary).
  \item The process at the node \fbox{$+$} can be implemented in the same way as the message processing at a generic check node of an LDPC code (binary or non-binary).
  \item The process at the node \fbox{$\Pi$} is the same as the original one, which interleaves or deinterleaves the input messages.
\end{itemize}

{\color{black}
Upon the arrival of the received vector $\bm{y}^{(t)}$~(corresponding to the sub-block $\bm{c}^{(t)}$) at time $t$, the SWD algorithm takes as inputs the \emph{a posterior} probabilities~(APPs) corresponding to $\bm{C}^{(t)}$ and uses the APPs corresponding to $\bm{C}^{(t-d)}, \cdots, \bm{C}^{(t)}$ to recover $\bm{u}^{(t-d)}$, where the computation of APPs
is similar to (\ref{eq:channelAPP}). After $\bm{u}^{(t-d)}$ is recovered, the decoder discards $\bm{y}^{(t-d)}$ and slides one layer of the normal graph to the ``right" to recover $\bm{u}^{(t-d+1)}$ with $\bm{y}^{(t+1)}$ received.
}

\section{Examples of BMST-RUN Codes}\label{sec:Examples}
\begin{table}[t]
\caption{Construction Examples of BMST-RUN Codes over AWGN Channels\label{tab:MemoryRequired}}
\centering
\begin{tabular}{p{0.95cm}p{0.15cm}p{0.9cm}cp{0.3cm}p{0.5cm}rrl}
  \hline
  \hline
  \scriptsize $\mathcal{A}$ & \scriptsize $\frac{P}{Q}$ & \scriptsize \scriptsize $\left(\frac{1}{N+1}, \frac{1}{N}\right)$ & \scriptsize $\alpha$ & \scriptsize $B$ & \scriptsize $p_{\rm target}$ & \scriptsize $\gamma_{\lim}$ (dB) & \scriptsize \color{black} $m$ & \scriptsize \color{black} $\varphi$$^{*}$\\

  \hline
  \scriptsize BPSK & \scriptsize $\frac{1}{8}$ & \scriptsize $\left(\frac{1}{9}, \frac{1}{8}\right)$ & \scriptsize $0$ & \scriptsize $1250$ & \scriptsize $10^{-5}$ & \scriptsize $-7.2$ & \scriptsize $11$ & \scriptsize $\varphi_0$ \\
  \scriptsize BPSK & \scriptsize $\frac{2}{8}$ & \scriptsize $\left(\frac{1}{5}, \frac{1}{4}\right)$ & \scriptsize $0$ & \scriptsize $1250$ & \scriptsize $10^{-5}$ & \scriptsize $-3.8$ & \scriptsize $10$ & \scriptsize $\varphi_0$ \\
  \scriptsize BPSK & \scriptsize $\frac{3}{8}$ & \scriptsize $\left(\frac{1}{3}, \frac{1}{2}\right)$ & \scriptsize $\frac{2}{3}$ & \scriptsize $1250$ & \scriptsize $10^{-5}$ & \scriptsize $-1.6$ & \scriptsize $11$ & \scriptsize $\varphi_0$\\
  \scriptsize BPSK & \scriptsize $\frac{4}{8}$ & \scriptsize $\left(\frac{1}{3}, \frac{1}{2}\right)$ & \scriptsize $0$ & \scriptsize $1250$ & \scriptsize $10^{-5}$ & \scriptsize $ 0.2$ & \scriptsize $8$ & \scriptsize $\varphi_0$\\
  \scriptsize BPSK & \scriptsize $\frac{5}{8}$ & \scriptsize $\left(\frac{1}{2}, 1\right)$ & \scriptsize $\frac{3}{5}$ & \scriptsize $1250$ & \scriptsize $10^{-5}$ & \scriptsize $1.8$ & \scriptsize $10$ & \scriptsize $\varphi_0$\\
  \scriptsize BPSK & \scriptsize $\frac{6}{8}$ & \scriptsize $\left(\frac{1}{2}, 1\right)$ & \scriptsize $\frac{1}{3}$ & \scriptsize $1250$ & \scriptsize $10^{-5}$ & \scriptsize $ 3.4$ & \scriptsize $7$ & \scriptsize $\varphi_0$\\
  \scriptsize BPSK & \scriptsize $\frac{7}{8}$ & \scriptsize $\left(\frac{1}{2}, 1\right)$ & \scriptsize $\frac{1}{7}$ & \scriptsize $1250$ & \scriptsize $10^{-5}$ & \scriptsize $ 5.3$ & \scriptsize $5$ & \scriptsize $\varphi_0$\\

  \hline
  \scriptsize $3$-PAM & \scriptsize $\frac{1}{7}$ & \scriptsize $\left(\frac{1}{8}, \frac{1}{7}\right)$ & \scriptsize $0$ & \scriptsize $300$ & \scriptsize $10^{-4}$ & \scriptsize $-4.3$ & \scriptsize $7$ & \scriptsize $\varphi_1$ \\
  \scriptsize $3$-PAM & \scriptsize $\frac{2}{7}$ & \scriptsize $\left(\frac{1}{4}, \frac{1}{3}\right)$ & \scriptsize $\frac{1}{2}$ & \scriptsize $300$ & \scriptsize $10^{-4}$ & \scriptsize $-0.5$ & \scriptsize $6$ & \scriptsize $\varphi_1$ \\
  \scriptsize $3$-PAM & \scriptsize $\frac{3}{7}$ & \scriptsize $\left(\frac{1}{3}, \frac{1}{2}\right)$ & \scriptsize $\frac{1}{3}$ & \scriptsize $300$ & \scriptsize $10^{-4}$ & \scriptsize $2.1$ & \scriptsize $6$ & \scriptsize $\varphi_1$ \\
  \scriptsize $3$-PAM & \scriptsize $\frac{4}{7}$ & \scriptsize $\left(\frac{1}{2}, 1\right)$ & \scriptsize $\frac{3}{4}$ & \scriptsize $300$ & \scriptsize $10^{-4}$ & \scriptsize $4.4$ & \scriptsize $6$ & \scriptsize $\varphi_1$ \\
  \scriptsize $3$-PAM & \scriptsize $\frac{5}{7}$ & \scriptsize $\left(\frac{1}{2}, 1\right)$ & \scriptsize $\frac{2}{5}$ & \scriptsize $300$ & \scriptsize $10^{-4}$ & \scriptsize $6.5$ & \scriptsize $5$ & \scriptsize $\varphi_1$ \\
  \scriptsize $3$-PAM & \scriptsize $\frac{6}{7}$ & \scriptsize $\left(\frac{1}{2}, 1\right)$ & \scriptsize $\frac{1}{6}$ & \scriptsize $300$ & \scriptsize $10^{-4}$ & \scriptsize $8.8$ & \scriptsize $3$ & \scriptsize $\varphi_1$ \\

  \hline
  \scriptsize $4$-PAM & \scriptsize $\frac{1}{7}$ & \scriptsize $\left(\frac{1}{8}, \frac{1}{7}\right)$ & \scriptsize $0$ & \scriptsize $200$ & \scriptsize $10^{-4}$ & \scriptsize $-3.1$ & \scriptsize $9$ & \scriptsize $\varphi_3$ \\
  \scriptsize $4$-PAM & \scriptsize $\frac{2}{7}$ & \scriptsize $\left(\frac{1}{4}, \frac{1}{3}\right)$ & \scriptsize $\frac{1}{2}$ & \scriptsize $200$ & \scriptsize $10^{-4}$ & \scriptsize $0.9$ & \scriptsize $8$ & \scriptsize $\varphi_3$ \\
  \scriptsize $4$-PAM & \scriptsize $\frac{3}{7}$ & \scriptsize $\left(\frac{1}{3}, \frac{1}{2}\right)$ & \scriptsize $\frac{1}{3}$ & \scriptsize $200$ & \scriptsize $10^{-4}$ & \scriptsize $3.8$ & \scriptsize $6$ & \scriptsize $\varphi_3$ \\
  \scriptsize $4$-PAM & \scriptsize $\frac{4}{7}$ & \scriptsize $\left(\frac{1}{2}, 1\right)$ & \scriptsize $\frac{3}{4}$ & \scriptsize $200$ & \scriptsize $10^{-4}$ & \scriptsize $6.3$ & \scriptsize $7$ & \scriptsize $\varphi_3$ \\
  \scriptsize $4$-PAM & \scriptsize $\frac{5}{7}$ & \scriptsize $\left(\frac{1}{2}, 1\right)$ & \scriptsize $\frac{2}{5}$ & \scriptsize $200$ & \scriptsize $10^{-4}$ & \scriptsize $8.7$ & \scriptsize $5$ & \scriptsize $\varphi_3$ \\
  \scriptsize $4$-PAM & \scriptsize $\frac{6}{7}$ & \scriptsize $\left(\frac{1}{2}, 1\right)$ & \scriptsize $\frac{1}{6}$ & \scriptsize $200$ & \scriptsize $10^{-4}$ & \scriptsize $11.2$ & \scriptsize $3$ & \scriptsize $\varphi_3$ \\

  \hline
  \scriptsize $8$-PSK & \scriptsize $\frac{1}{5}$ & \scriptsize $\left(\frac{1}{6}, \frac{1}{5}\right)$ & \scriptsize $0$ & \scriptsize $150$ & \scriptsize $10^{-4}$ & \scriptsize $-2.8$ & \scriptsize $6$ & \scriptsize $\varphi_5$ \\
  \scriptsize $8$-PSK & \scriptsize $\frac{2}{5}$ & \scriptsize $\left(\frac{1}{3}, \frac{1}{2}\right)$ & \scriptsize $\frac{1}{2}$ & \scriptsize $150$ & \scriptsize $10^{-4}$ & \scriptsize $1.3$ & \scriptsize $6$ & \scriptsize $\varphi_6$ \\
  \scriptsize $8$-PSK & \scriptsize $\frac{3}{5}$ & \scriptsize $\left(\frac{1}{2}, 1\right)$ & \scriptsize $\frac{2}{3}$ & \scriptsize $150$ & \scriptsize $10^{-4}$ & \scriptsize $4.7$ & \scriptsize $6$ & \scriptsize $\varphi_6$ \\
  \scriptsize $8$-PSK & \scriptsize $\frac{4}{5}$ & \scriptsize $\left(\frac{1}{2}, 1\right)$ & \scriptsize $\frac{1}{4}$ & \scriptsize $150$ & \scriptsize $10^{-4}$ & \scriptsize $8.1$ & \scriptsize $4$ & \scriptsize $\varphi_6$ \\
  \hline
  \scriptsize $16$-QAM & \scriptsize $\frac{239}{255}$ & \scriptsize $\left(\frac{1}{2}, 1\right)$ & \scriptsize $\frac{16}{239}$ & \scriptsize $4$ & \scriptsize $10^{-3}$ & \scriptsize $12.7$ & \scriptsize $2$ & \scriptsize $\varphi_7$ \\
  \hline
  \scriptsize uniformly spaced $16$-PAM & \scriptsize $\frac{1}{2}$ & \scriptsize $\left(\frac{1}{3}, \frac{1}{2}\right)$ & \scriptsize $0$ & \scriptsize $250$ & \scriptsize $10^{-3}$ & \scriptsize $12.5$ & \scriptsize $5$ & \scriptsize $\varphi_8$ \\
  \scriptsize non-uniformly spaced $16$-PAM~\cite{Sun93} & \scriptsize $\frac{1}{2}$ & \scriptsize $\left(\frac{1}{3}, \frac{1}{2}\right)$ & \scriptsize $0$ & \scriptsize $250$ & \scriptsize $10^{-3}$ & \scriptsize $12.0$ & \scriptsize $5$ & \scriptsize $\varphi_9$ \\
  \hline
\end{tabular}
{\color{black}
\begin{tablenotes}
\footnotesize
\item{*} The mappings in this table are the same as those specified in Fig.~\ref{fig:SignalSetsAndMappings}.
     Notice that the shaping gain of the non-uniformly spaced $16$-PAM is about $0.5$~dB.

\end{tablenotes}
}
\end{table}
In this section, we present simulation results of several BMST-RUN codes over AWGN channels {\color{black}and Rayleigh flat fading channels}, where code parameters are shown in Table~\ref{tab:MemoryRequired}.
For all simulations, the encoder terminates every $L=1000$ sub-blocks and the decoder executes the SWD algorithm with a maximum iteration number $18$. Without specification, the decoding delay $d$ of the SWD algorithm is set to be $3m$.

\subsection{BMST-RUN Codes with One-Dimensional Signal Sets}\label{subsec:OneDim}
\begin{figure}[t]
\centering
  \includegraphics[width=\figwidth]{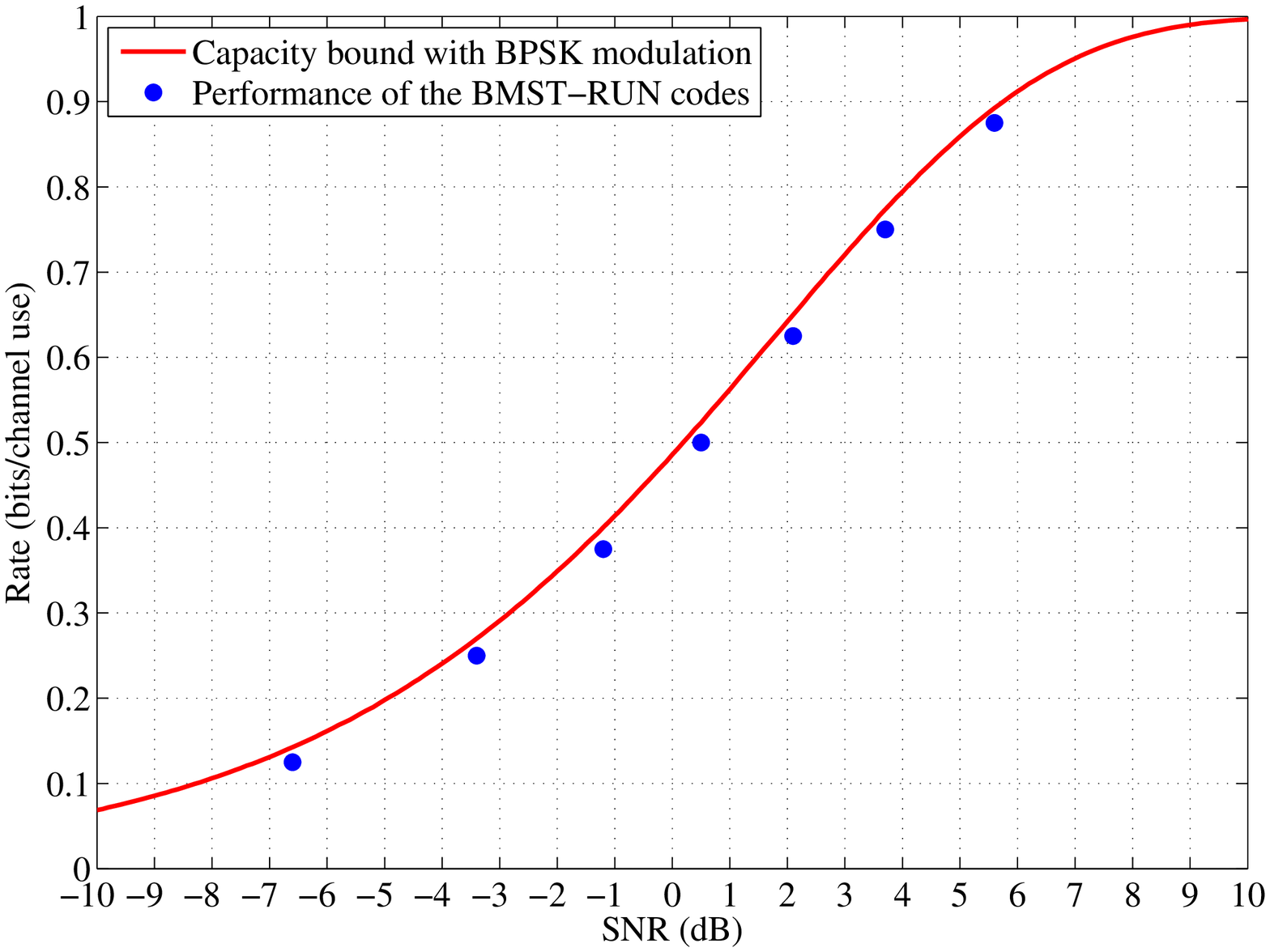} \\
  \vspacecaption
  \caption{The required SNRs to achieve the SER of $10^{-5}$ for the BMST-RUN codes with the codes $\mathcal{C}_{\rm RUN}[Q,P]^{1250}(\frac{P}{Q}=\frac{1}{8}, \cdots, \frac{7}{8})$ as basic codes defined with BPSK modulation.}
  \label{fig:BPSKcapacity}
  \vspacefigure
\end{figure}
Consider BMST-RUN codes of rates $\frac{K}{8}(K=1,\cdots,7)$ defined with BPSK modulation to approach the Shannon limits at the SER of $10^{-5}$. Fig.~\ref{fig:BPSKcapacity} shows the required SNRs for the BMST-RUN codes to achieve the SER of $10^{-5}$. Also shown in Fig.~\ref{fig:BPSKcapacity} is the channel capacity curve with i.u.d. inputs. It can be seen that the gaps between the required SNRs and the Shannon limits are within $1$~dB for all considered rates.

{\color{black}
\begin{figure}[t]
\centering
  \includegraphics[width=\figwidth]{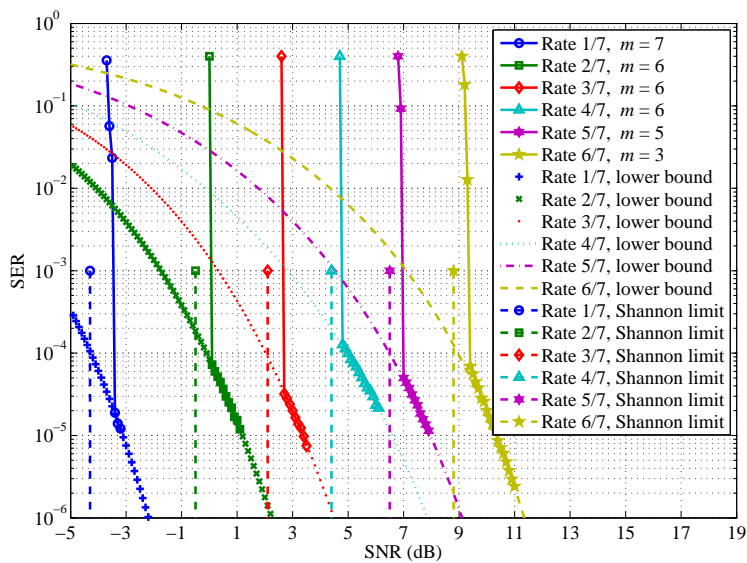}\\
  \vspacecaption
  \caption{Performances of the BMST-RUN codes with the codes $\mathcal{C}_{\rm RUN}[Q,P]^{300}$ $(\frac{P}{Q}=\frac{1}{7}, \cdots, \frac{6}{7})$ as basic codes defined with $3$-PAM.}
  \label{fig:3AMsnrser}
  \vspacefigure
\end{figure}
Consider BMST-RUN codes of rates $\frac{K}{7}(K\!=\!1,\!\cdots\!,\!6)$ defined with $3$-PAM to approach the Shannon limits at the SER of $10^{-4}$. Fig.~\ref{fig:3AMsnrser} shows the SER performance curves for all codes together with their lower bounds and the corresponding Shannon limits. We can see that the performance curves match well with the corresponding lower bounds for all codes in the high SNR region. In addition, all codes have an SER lower than $10^{-4}$ at the SNR within $1$ dB away from the corresponding Shannon limits, which is similar to the BPSK modulation case.}

\begin{figure}[t]
\centering
  \includegraphics[width=\figwidth]{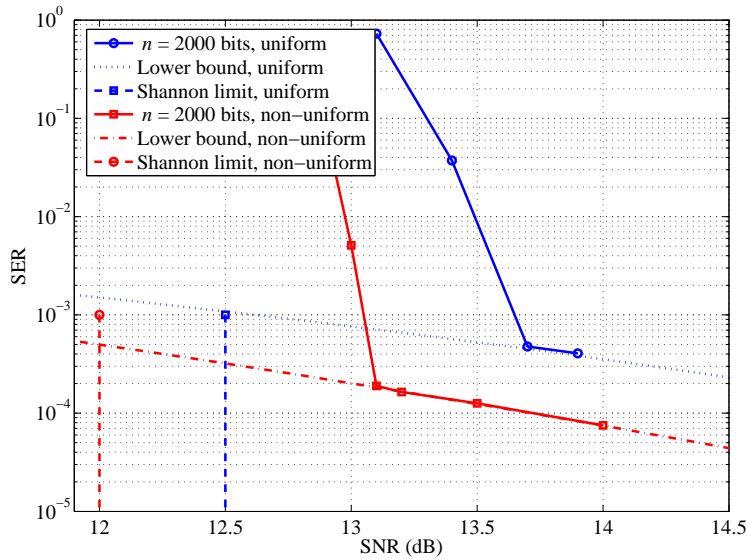}\\
  \vspacecaption
  \caption{\color{black}Comparison of the BMST-RUN code with the code $\mathcal{C}_{\rm RUN}[2,1]^{250}$ as the basic code defined with two distinct $16$-PAM constellations under the mapping $\varphi_8$ and $\varphi_9$ in Fig.~\ref{fig:SignalSetsAndMappings}.}
  \label{fig:16AMsnrser}
  \vspacefigure
\end{figure}
{\color{black}Consider a rate-$\frac{1}{2}$ BMST-RUN code of memory $5$ defined
over two distinct $16$-PAM constellations, where one consists of uniformly spaced signal points (under the mapping $\varphi_8$ in Fig.~\ref{fig:SignalSetsAndMappings}) and the other consists of non-uniformly spaced signal points (under the mapping $\varphi_9$ in Fig.~\ref{fig:SignalSetsAndMappings}) as designed in~\cite{Sun93}.
The SER performance curves with a decoding delay $d=15$ together with the lower bounds and the Shannon limits are shown in Fig.~\ref{fig:16AMsnrser}. From the figure, we can see that the BMST-RUN code has an SER lower than $10^{-3}$ at the SNR about $1.0$ away from their respective Shannon limits for both uniformly spaced signal points and non-uniformly spaced signal points. In addition,
the BMST-RUN code with non-uniformly spaced signal points performs about $0.5$~dB better than that with uniformly spaced signal points and also has a lower error floor.}

\subsection{BMST-RUN Codes with Two-Dimensional Signal Sets}
\begin{figure}[t]
\centering
  \includegraphics[width=\figwidth]{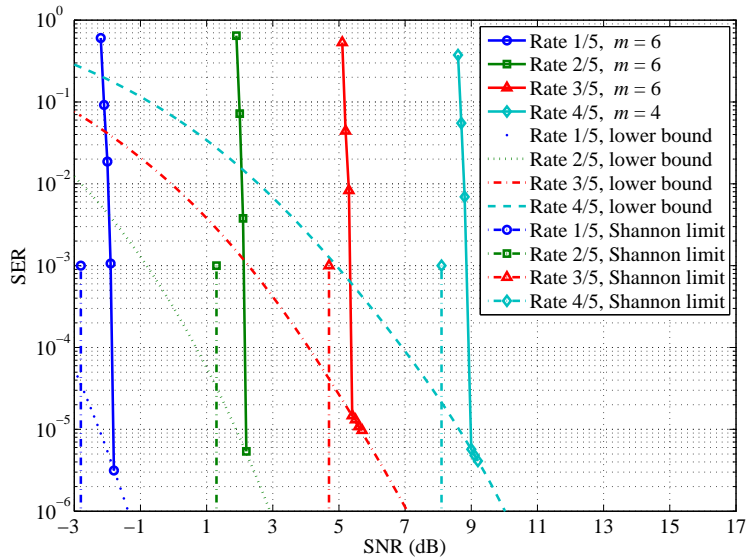}\\
  \vspacecaption
  \caption{Performances of the BMST-RUN codes with the codes $\mathcal{C}_{\!\rm RUN\!}[Q\!,\!P]^{\!150\!}(\!\frac{P}{Q}\!=\!\frac{1}{5}, \!\cdots\!, \!\frac{4}{5}\!)$ as basic codes defined with $8$-PSK modulation.}
  \label{fig:8PSKsnrser}
  \vspacefigure
\end{figure}
Consider BMST-RUN codes of rates $\frac{K}{5}(K\!=\!1,\!\cdots\!,\!4)$ defined with $8$-PSK modulation to approach the Shannon limits at the SER of $10^{-4}$. Fig.~\ref{fig:8PSKsnrser} shows the SER performance curves for all codes together with their lower bounds and the corresponding Shannon limits.

{\color{black}
\begin{figure}[t]
\centering
  \includegraphics[width=\figwidth]{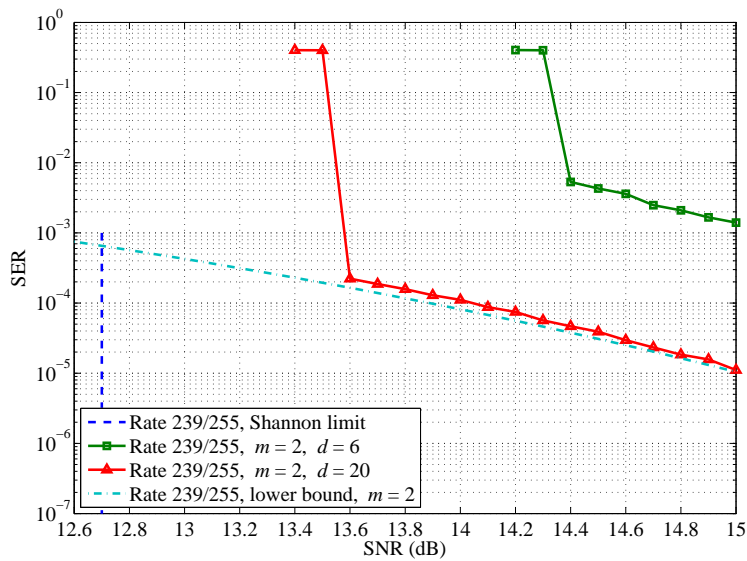}\\
  \vspacecaption
  \caption{\color{black}Performance of the BMST-RUN code with the code $\mathcal{C}_{\rm RUN}[255,239]^{4}$ as the basic code defined with $16$-QAM, where the mapping is $\varphi_7$ in Fig.~\ref{fig:SignalSetsAndMappings}.}
  \label{fig:16QAMsnrser}
  \vspacefigure
\end{figure}
Consider a BMST-RUN code of rate $\frac{239}{255}$ defined with $16$-QAM~(under the mapping $\varphi_7$ in Fig.~\ref{fig:SignalSetsAndMappings}) to approach the Shannon limit at the SER of $10^{-3}$, where an encoding memory $m=2$ is required. The SER performance curves with decoding delays $d=6$ and $20$ together with the lower bound and the Shannon limit are shown in Fig.~\ref{fig:16QAMsnrser}. Since a large fraction of information symbols~($\frac{223}{239}$) are uncoded in the basic code, a large decoding delay $d=10m=20$ is required to approach the lower bound. With the decoding delay $d=20$, the BMST-RUN code achieves the SER of $10^{-3}$ at the SNR about $1$~dB away from the Shannon limit.

{\color{black}From the above two examples, we can see that BMST codes with two-dimensional signal constellations behave similarly as they do with one-dimensional signal constellations.
}
}

{\color{black}
\subsection{Comparison with BMST-BICM}\label{subsec:Comparison}
\begin{figure}[t]
\centering
  \includegraphics[width=\figwidth]{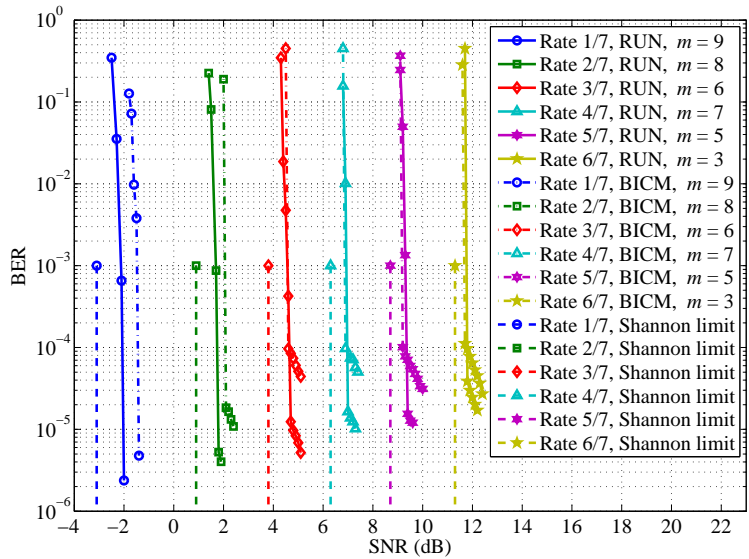}\\
  \vspacecaption
  \caption{\color{black}Performance of the BMST-RUN codes with the codes $\mathcal{C}_{\rm RUN}[7,K]^{200}(K\!=\!1,\!\cdots\!,\!6)$ over the modulo-$4$ group and the BMST-BICM scheme with the codes $\mathcal{C}_{\rm RUN}[7,K]^{400}(K\!=\!1,\!\cdots\!,\!6)$ over $\mathbb{F}_2$ as basic codes, where both schemes are under $4$-PAM with the mapping $\varphi_3$ in Fig.~\ref{fig:SignalSetsAndMappings}.}
  \label{fig:4AMsnrser}
  \vspacefigure
\end{figure}
\begin{figure}[t]
\centering
  \includegraphics[width=\figwidth]{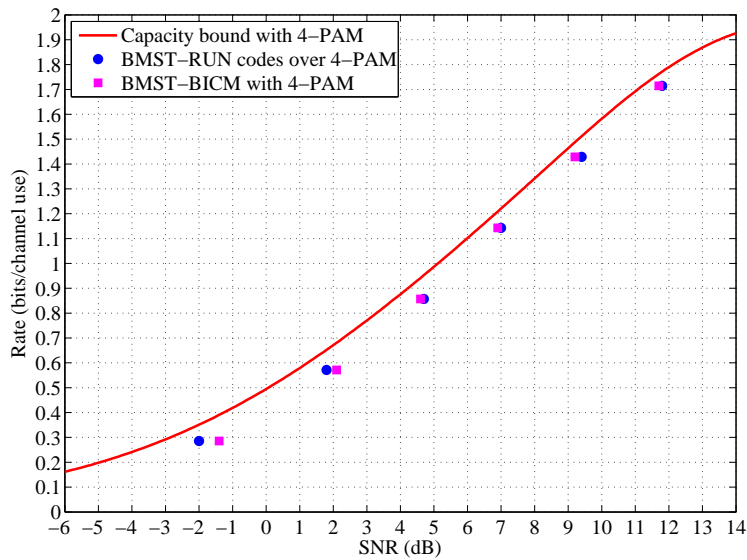}\\
  \vspacecaption
  \caption{\color{black}The required SNRs to achieve the BER of $10^{-4}$ {\color{black}over AWGN channels} for the BMST-RUN codes with the codes $\mathcal{C}_{\rm RUN}[7,K]^{200}(K\!=\!1,\!\cdots\!,\!6)$ over the modulo-$4$ group and the BMST-BICM scheme with the codes $\mathcal{C}_{\rm RUN}[7,K]^{400}(K\!=\!1,\!\cdots\!,\!6)$ over $\mathbb{F}_2$ as basic codes, where both schemes are under $4$-PAM with the mapping $\varphi_3$ in Fig.~\ref{fig:SignalSetsAndMappings}.}
  \label{fig:4AMsnrcapacity}
  \vspacefigure
\end{figure}
\color{black}
The examples in the previous subsections suggest that the proposed construction is effective for a wide range of code rates and signal sets. Also, the SWD algorithm is near-optimal in the high SNR region.
Since binary BMST codes also have such behaviors and can be combined with different signal sets~\cite{Liang14}, we need clarify the advantage of BMST-RUN codes over groups. Some advantages have been mentioned in the Introduction. In this subsection, we will show that the BMST-RUN codes can perform better than the BMST-BICM scheme.


\color{black}

\color{black}
To make a fair comparison, we have the following settings.
\begin{itemize}
  \item For the BMST-BICM scheme, the basic codes are the RUN codes $[7,K]^{400}(K\!=\!1,\!\cdots\!,\!6)$ over $\mathbb{F}_2$, while for the BMST-RUN codes, the basic codes are the RUN codes $[7,K]^{200}(K\!=\!1,\!\cdots\!,\!6)$ over the modulo-$4$ group. Such setting ensures that both schemes have the same sub-block length {\color{black}$2800$ in bits}.

  \item Both the BMST-RUN codes and the BMST-BICM scheme use the $4$-PAM with the mapping $\varphi_3$ in Fig.~\ref{fig:SignalSetsAndMappings}.

  \item For a specific code rate, the BMST-BICM scheme has the same encoding memory and the same decoding delay as the BMST-RUN code. The encoding memories are presented in Table~\ref{tab:MemoryRequired}, while the decoding delay is set to be $3m$ for an encoding memory $m$.
\end{itemize}

Since the performance of the BMST-BICM scheme can not be measured in SER, we compare the performance in BER.
Fig.~\ref{fig:4AMsnrser} shows the BER performance curves for both the BMST-RUN codes~(denoted as ``RUN") and the BMST-BICM scheme~(denoted as ``BICM") together with the Shannon limits.
Fig.~\ref{fig:4AMsnrcapacity} shows the required SNRs to achieve the BER of $10^{-4}$ for both the BMST-RUN codes and the BMST-BICM scheme together with capacity curve of $4$-PAM under i.u.d. inputs.
From these two figures, we have the following observations.
\begin{itemize}
  \item With the same encoding memory and decoding delay, the BMST-RUN codes achieve a lower BER than the BMST-BICM scheme for all considered code rates.
  \item The BMST-RUN codes perform better than the BMST-BICM scheme in the lower code rate region and have a similar performance as the BMST-BICM scheme in the high code rate region.
\end{itemize}
}

\subsection{\color{black}BMST-RUN Codes over Rayleigh Channels}\label{subsec:Rayleigh}
\begin{figure}[t]
\centering
  \includegraphics[width=\figwidth]{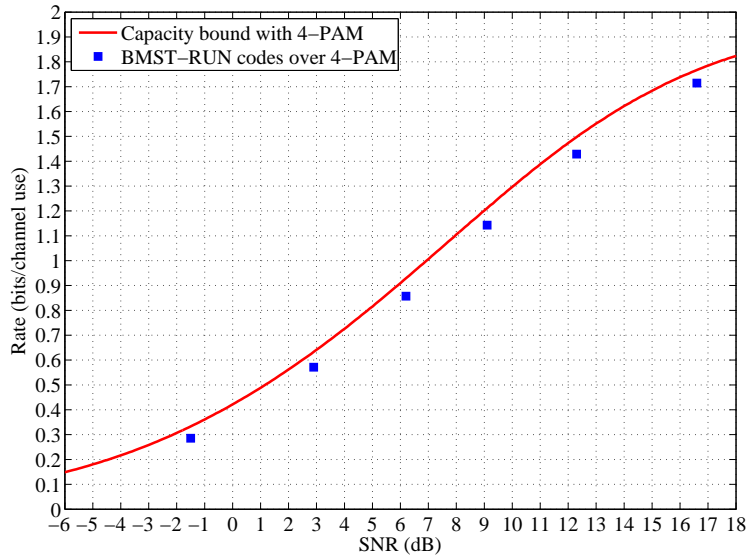} \\
  \vspacecaption
  \caption{\color{black}The required SNRs to achieve the SER of $10^{-4}$ for the BMST-RUN codes with the codes $\mathcal{C}_{\rm RUN}[Q,P]^{200}(\frac{P}{Q}=\frac{1}{7}, \cdots, \frac{6}{7})$ as basic codes defined with 4-PAM modulation~(under the mapping $\varphi_3$ in Fig.~\ref{fig:SignalSetsAndMappings}) over Rayleigh flat fading channels.}
  \label{fig:4AMRayleighcapacity}
  \vspacefigure
\end{figure}
{\color{black}It has been shown that BMST-RUN codes perform well over AWGN channels and are comparable to binary BMST codes with BICM. More interestingly and importantly, BMST construction is also applicable to other ergodic channels. Here,  we give an example for fading channels as an evidence.}

{\color{black}
Consider BMST-RUN codes of rates $\frac{K}{7}(K=1,\cdots,6)$ defined with 4-PAM modulation~(under the mapping $\varphi_3$ in Fig.~\ref{fig:SignalSetsAndMappings}) over Rayleigh flat fading channels. To approach the Shannon limits at the SER of $10^{-4}$, \color{black}the required encoding memories for rates $\frac{1}{7},\frac{2}{7},\frac{3}{7},\frac{4}{7},\frac{5}{7},$ and $\frac{6}{7}$ are $7,7,6,7,5,$ and $4$, respectively. Fig.~\ref{fig:4AMRayleighcapacity} shows the required SNRs for the BMST-RUN codes to achieve the SER of $10^{-4}$. Also shown in Fig.~\ref{fig:4AMRayleighcapacity} is the channel capacity curve with i.u.d. inputs. It can be seen that the gaps between the required SNRs and the Shannon limits are about $1$~dB for all rates, which is similar to the case for AWGN channels.}

\section{Conclusions}\label{sec:Conclusion}
In this paper, by combining the block Markov superposition transmission~(BMST) with {\color{black}the RUN codes over groups,} we have proposed a simple scheme called BMST-RUN codes to approach the Shannon limits at any target symbol-error-rate~(SER) with any given (rational) rate over any alphabet (of moderate size). We have also derived the genie-aided lower bound for the BMST-RUN codes. Simulation results have shown that the BMST-RUN codes have a similar behavior to the binary BMST codes and have good performance for a wide range of code rates over {\color{black}both} AWGN channels {\color{black}and Rayleigh flat fading channels}. {\color{black}Compared with the BMST with bit-interleaved coded modulation~(BMST-BICM) scheme, the BMST-RUN codes are more flexible, which {\color{black}can be combined} with signal sets of any size. In addition, with the same encoding memory, the BMST-RUN codes have a better performance than the BMST-BICM scheme under the same decoding latency.}
\section*{Acknowledgment}
The authors wish to thank Mr. Kechao Huang and Mr. Jinshun Zhu for useful discussions.


\begin{thebibliography}{10}
\providecommand{\url}[1]{#1}
\csname url@samestyle\endcsname
\providecommand{\newblock}{\relax}
\providecommand{\bibinfo}[2]{#2}
\providecommand{\BIBentrySTDinterwordspacing}{\spaceskip=0pt\relax}
\providecommand{\BIBentryALTinterwordstretchfactor}{4}
\providecommand{\BIBentryALTinterwordspacing}{\spaceskip=\fontdimen2\font plus
\BIBentryALTinterwordstretchfactor\fontdimen3\font minus
  \fontdimen4\font\relax}
\providecommand{\BIBforeignlanguage}[2]{{%
\expandafter\ifx\csname l@#1\endcsname\relax
\typeout{** WARNING: IEEEtran.bst: No hyphenation pattern has been}%
\typeout{** loaded for the language `#1'. Using the pattern for}%
\typeout{** the default language instead.}%
\else
\language=\csname l@#1\endcsname
\fi
#2}}
\providecommand{\BIBdecl}{\relax}
\BIBdecl

\bibitem{Berrou93}
C.~Berrou, A.~Glavieux, and P.~Thitimajshima, ``{Near Shannon limit
  error-correcting coding and decoding: Turbo codes},'' in \emph{Proc. Int.
  Conf. Commun.}, Geneva, Switzerland, May 1993, pp. 1064--1070.

\bibitem{Gallager63}
R.~G. Gallager, \emph{{Low-Density Parity-Check Codes}}.\hskip 1em plus 0.5em
  minus 0.4em\relax Cambridge, MA, USA: MIT Press, 1963.

\bibitem{Felstrom99}
A.~Jim\'{e}nez-Felstr\"{o}m and K.~S. Zigangirov, ``{Time-varying periodic
  convolutional codes with low-density parity-check matrix},'' \emph{IEEE
  Trans. Inf. Theory}, vol.~45, no.~6, pp. 2181--2191, Sep. 1999.

\bibitem{Kudekar11}
S.~Kudekar, T.~J. Richardson, and R.~L. Urbanke, ``{Threshold saturation via
  spatial coupling: Why convolutional LDPC ensembles perform so well over the
  BEC},'' \emph{IEEE Trans. Inf. Theory}, vol.~57, no.~2, pp. 803--834, Feb.
  2011.

\bibitem{Feltstrom09}
A.~Jim\'{e}nez-Felstr\"{o}m, D.~Truhachev, M.~Lentmaier, and K.~S. Zigangirov,
  ``Braided block codes,'' \emph{IEEE Trans. Inf. Theory}, vol.~55, no.~6, pp.
  2640--2658, Jun. 2009.

\bibitem{Smith12}
B.~P. Smith, A.~Farhood, A.~Hunt, F.~R. Kschischang, and J.~Lodge, ``{Staircase
  codes: FEC for 100 Gb/s OTN},'' \emph{J. Lightw. Technol.}, vol.~30, no.~1,
  pp. 110--117, Jan. 2012.

\bibitem{Moloudi14}
S.~Moloudi, M.~Lentmaier, and A.~Graell~i Amatz, ``Spatially coupled turbo
  codes,'' in \emph{Proc. 8th Int. Symp. Turbo Codes}, Bremen, Germany, Aug.
  2014, pp. 82--86.

\bibitem{Ma13}
X.~Ma, C.~Liang, K.~Huang, and Q.~Zhuang, ``{Obtaining extra coding gain for
  short codes by block Markov superposition transmission},'' in \emph{Proc.
  IEEE Int. Symp. Inf. Theory}, Istanbul, Turkey, Jul. 2013, pp. 2054--2058.

\bibitem{Ma15}
------, ``{Block Markov superposition transmission: Construction of big
  convolutional codes from short codes},'' \emph{IEEE Trans. Inf. Theory},
  vol.~61, no.~6, pp. 3150--3163, Jun. 2015.

\bibitem{Liang14c}
C.~Liang, X.~Ma, Q.~Zhuang, and B.~Bai, ``{Spatial coupling of generator
  matrices: A general approach to design good codes at a target BER},''
  \emph{IEEE Trans. Commun.}, vol.~62, no.~12, pp. 4211--4219, Dec. 2014.

\bibitem{Hu14}
J.~Hu, C.~Liang, X.~Ma, and B.~Bai, ``A new class of multiple-rate codes based
  on block {M}arkov superposition transmission,'' in \emph{Proc. 3rd Int.
  Workshop High Mobility Wirel. Commun.}, Beijing, China, Nov. 2014, pp.
  109--114.

\bibitem{Liang15}
C.~Liang, J.~Hu, X.~Ma, and B.~Bai, ``{A new class of multiple-rate codes based
  on block Markov superposition transmission},'' \emph{IEEE Trans. Signal
  Process.}, vol.~63, no.~16, pp. 4236--4244, Aug. 15 2015.

\bibitem{Hu15}
J.~Hu, X.~Ma, and C.~Liang, ``Block {M}arkov superposition transmission of
  repetition and single-parity-check codes,'' \emph{IEEE Commun. Lett.},
  vol.~19, no.~2, pp. 131--134, Feb. 2015.

\bibitem{Liang14}
C.~Liang, K.~Huang, X.~Ma, and B.~Bai, ``{Block Markov superposition
  transmission with bit-interleaved coded modulation},'' \emph{IEEE Commun.
  Lett.}, vol.~18, no.~3, pp. 397--400, Mar. 2014.

\bibitem{Yang14}
Z.~Yang, C.~Liang, X.~Xu, and X.~Ma, ``Block {M}arkov superposition
  transmission with spatial modulation,'' \emph{IEEE Wirel. Commun. Lett.},
  vol.~3, no.~6, pp. 565--568, Dec. 2014.

\bibitem{Liu15}
X.~Liu, C.~Liang, and X.~Ma, ``Block {M}arkov superposition transmission of
  convolutional codes with minimum shift keying signalling,'' \emph{IET
  Commun.}, vol.~9, no.~1, pp. 71--77, Jan. 2015.

\bibitem{Xu15}
X.~Xu, C.~Wang, Y.~Zhu, X.~Ma, and X.~Zhang, ``Block {M}arkov superposition
  transmission of short codes for indoor visible light communications,''
  \emph{IEEE Commun. Lett.}, vol.~19, no.~3, pp. 359--362, Mar. 2015.

\bibitem{Sun93}
F.-W. Sun and H.~C.~A. van Tilborg, ``Approaching capacity by equiprobable
  signaling on the {G}aussian channel,'' \emph{IEEE Trans. Inf. Theory},
  vol.~39, no.~5, pp. 1714--1716, Sep. 1993.

\bibitem{Ma04}
X.~Ma and L.~Ping, ``Coded modulation using superimposed binary codes,''
  \emph{IEEE Trans. Inf. Theory}, vol.~50, no.~12, pp. 3331--3343, Dec. 2004.

\bibitem{Yang12}
Z.~Yang, S.~Zhao, X.~Ma, and B.~Bai, ``A new joint source-channel coding scheme
  based on nested lattice codes,'' \emph{IEEE Commun. Lett.}, vol.~16, no.~5,
  pp. 730--733, May 2012.

\bibitem{Lentmaier10}
M.~Lentmaier, A.~Sridharan, D.~J. Costello, Jr., and K.~S. Zigangirov,
  ``{Iterative decoding threshold analysis for LDPC convolutional codes},''
  \emph{IEEE Trans. Inf. Theory}, vol.~56, no.~10, pp. 5274--5289, Oct. 2010.

\bibitem{Iyengar12}
A.~R. Iyengar, M.~Papaleo, P.~H. Siegel, J.~K. Wolf, A.~Vanelli-Coralli, and
  G.~E. Corazza, ``Windowed decoding of protograph-based {LDPC} convolutional
  codes over erasure channels,'' \emph{IEEE Trans. Inf. Theory}, vol.~58,
  no.~4, pp. 2303--2320, Apr. 2012.

\bibitem{Iyengar13}
A.~R. Iyengar, P.~H. Siegel, R.~L. Urbanke, and J.~K. Wolf, ``Windowed decoding
  of spatially coupled codes,'' \emph{IEEE Trans. Inf. Theory}, vol.~59, no.~4,
  pp. 2277--2292, Apr. 2013.

\bibitem{Huang15}
K.~Huang, X.~Ma, and D.~J. Costello, Jr., ``{EXIT} chart analysis of block
  {M}arkov superposition transmission of short codes,'' in \emph{Proc. IEEE
  Int. Symp. Inf. Theory}, Hong Kong, China, Jun. 2015.

\end{thebibliography}


\balance

\end{document}